\documentclass[showpacs,amssymb,floatfix,pre,aps,notitlepage]{revtex4-1}
\usepackage{amsmath}
\usepackage{amsfonts}
\usepackage{graphicx} 
\usepackage{placeins}
\usepackage{verbatim}
\usepackage{booktabs}
\usepackage{color}
\newcommand{\beq}{\begin{equation}}
\newcommand{\eeq}{\end{equation}}
\newcommand{\beqa}{\begin{eqnarray}}
\newcommand{\eeqa}{\end{eqnarray}}

\DeclareMathOperator{\bv}{\bf v }
\DeclareMathOperator*{\bG}{\bf N }
\DeclareMathOperator*{\too}{\to }

\DeclareMathOperator*{\erf}{erf }

\begin{document}
\title{Field induced stationary state for an accelerated tracer in a bath}
\author{Matthieu Barbier}
\author{Emmanuel Trizac}
\affiliation{Laboratoire de Physique Th\'eorique et Mod\`eles Statistiques, 
UMR CNRS 8626, Universit\'e Paris-Sud, 91405 Orsay, France}

\begin{abstract}
Our interest goes to the behavior of a tracer particle, accelerated 
by a constant and uniform external field, when the energy injected by the 
field is redistributed through collision to a bath of unaccelerated particles.
A non equilibrium steady state is thereby reached. Solutions of
a generalized Boltzmann-Lorentz equation are analyzed analytically, 
in a versatile framework that embeds the majority of tracer-bath interactions
discussed in the literature. These results --mostly derived
for a one dimensional system-- are successfully confronted
to those of three independent numerical simulation methods: 
a direct iterative solution, Gillespie algorithm, and the 
Direct Simulation Monte Carlo technique. We work out the diffusion properties
as well as the velocity
tails: large $v$, and either large $-v$, or $v$ in the vicinity of
its lower cutoff whenever the velocity distribution is bounded
from below. Particular emphasis is put on the cold bath limit, with 
scatterers at rest, which plays a special role in our model.
\end{abstract}


\maketitle

\section{Introduction}
\label{sec:intro}

The propagation of a single test particle (or "tracer") with mass $m_t$ accelerated by an external field $g$ in a stationary bath of particles with mass $m_b$ was first studied by Lorentz in 1905 \cite{Lorentz} as a kinetic approach of conduction in a metal. In this model, an electric field accelerates non-interacting electrons, which diffuse among a random configuration of static obstacles - the atoms - acting as infinitely massive scatterers ($m_t/m_b\to 0$), with a simple hard sphere electron-atom collision rule. Lorentz assumed the velocity distribution of the electrons to obey a Maxwell-Boltzmann law with a small correction, and was able to give an estimate of the conductivity of metals from microscopic parameters.
It was later shown that this perturbed Maxwell-Boltzmann distribution cannot be a stationary solution of the problem \cite{Pias79}: { the constant acceleration confers an ever-increasing amount of energy to the electron, which cannot be dissipated in the described collision process. Its velocity variance consequently diverges over time, while the mean velocity (and hence the current) stays finite and close to the Lorentz result on a certain time scale studied in \cite{Olaus,Pias93} then vanishes at long times}. However, a stationary solution may exist if the tracer is allowed to transfer some energy to the surrounding medium, for instance {when collisions are dissipative \cite{Martin07}} or with a finite mass ratio $m_t/m_b$. 

The case $m_b=m_t$ was therefore subsequently considered \cite{Pias83}, as an extension of R{\'e}sibois' field-less "self-diffusion problem" \cite{Resi78}, wherein one studies the diffusive properties of identical gas particles at equilibrium by following a single unaccelerated tracer. 
The bulk of ulterior developments on this modified Lorentz problem focused 
on asymptotic and relaxation properties \cite{Pias86,Gervois,Eder88,Martin99,Soto,Alas10}.
Recently, the model was used as a paradigm for the study of microscopic entropy production and fluctuation relations in non equilibrium systems \cite{Grad11,Puglisi11}, and to discuss the force-velocity 
relation \cite{Fiege} as measured in experiments \cite{Dauchot}.

In what follows, we will mostly concentrate on the one dimensional version of the model,
the several variants of which addressed in the literature differ first by the choice of different
bath velocity statistics $\Phi(v)$ (at rest with $\Phi(v)=\delta(v)$  \cite{Pias83}, dichotomous \cite{Pias86,Eder88,Soto} or Gaussian \cite{Gervois,Eder88,Alas10}) and second by the collision kernel 
accounting for grain interactions, that can be indexed by a continuous parameter
$\nu$ (hard rods with $\nu=1$ in most earlier works, 
but also $\nu=0$ for Maxwellian rods \cite{Maxwell} or the so-called "very hard rods" having 
$\nu=2$ \cite{Alas10}). Moreover, in addition to mass ratio and field strength $g$
that play a key role, collisions between the tracer and bath particles can be inelastic
\cite{Brill,Ernst06}. The ensuing parameter space is therefore large, with
associated rich behavior. At the Boltzmann equation level, our goal 
is here to provide a comprehensive view of the tracer stationary
velocity distribution function, that in all generality will be denoted 
$f_\nu(v,g,a|\Phi)$:
it depends on three parameters ($\nu$, $g$ and a
number $a$ that lumps mass ratio and dissipation in a single constant \cite{Martin99,Santos06}),
and is furthermore a functional of bath velocity distribution $\Phi(v)$. The reason for considering 
an arbitrary bath distribution lies in the wide spectrum of non-equilibrium
steady states that can be achieved upon driving granular gases with different energy 
injection mechanisms \cite{Montanero,Ernst06}; Stretched exponentials $\Phi(v) \sim \exp[-|v|^\mu/\mu]$
are often encountered here. A tracer in an evolving (e.g. freely cooling) bath may also be of interest \cite{kim2001statistical}, assuming that it can reach a state $f_\nu$ with no time dependence except through $\Phi(v,t)$.

Before rationalizing the behavior of $f_\nu$, the model will be detailed in section 
\ref{sec:model}, where a convenient integral form will be derived
from the Boltzmann equation. This provides the basis for an efficient
numerical resolution algorithm, that will be described in section 
\ref{num-methods}, together with the more versatile Gillespie and Direct
Simulation Monte Carlo techniques. Thereby equipped with three different numerical schemes, 
we will put to the test in section 
\ref{gen-bath} analytical results to be derived for the velocity tails and diffusive properties,
for an arbitrary bath distribution. The cold bath setting will be put forward 
and solved in Section \ref{Coldbath} as 
a relevant model onto which more general situations can be mapped,
while particular attention will be paid in Section \ref{non-negative} to the low velocity limit 
for bounded bath distributions. Finally, our conclusions will 
be presented in Section \ref{conclusion}.

\section{The model, its integral reformulation, and numerical resolution}
\label{sec:model}

\subsection{Statement of the problem}
\label{sub:stat}
We consider the one dimensional motion of a single particle of mass $m_t$, accelerated by a constant force $g>0$ through a bath. 
The bath is made up of unaccelerated particles of mass $m_b$, having a stationary velocity distribution function $\Phi(v)$ with characteristic velocity $v_b$. {This velocity scale is used for the nondimensionalization of both the velocity of the tracer $v = \hat v / v_b$ and the force $g= \hat g / \rho v_b^2$ where $\rho$ is the linear mass density in the bath, $\hat v$ and $\hat g$ being the corresponding variables with physical dimensions.}

The time dependent tracer velocity distribution is indexed by a parameter $\nu$ that specifies the type 
of tracer-bath interactions, and is denoted $F_{\nu}(v,t)$. The dependence on other parameters
and on $\Phi$ is left implicit unless necessary.

The dynamics under study is governed by the linear Boltzmann (or Boltzmann-Lorentz) equation 
\beq \dfrac{\partial F_{\nu}(v,t)}{\partial t} +g \dfrac{\partial F_{\nu}(v,t)}{\partial v} =  J[F_{\nu}(t), \Phi](v)  \label{FBoltz} \eeq
Here, $J$ is the collision operator 
\beq  J[\psi,\phi](v) =\int_{-\infty}^{+\infty} dv_1 dv_2 \, \psi(v_1) \,\phi(v_2) \,|v_1-v_2|^\nu \, \left[\delta(v_1'-v) - \delta(v_1-v) \right] 
\label{eq:nucoll}
\eeq 
where the post collisional velocity $v'_1$ of a binary encounter $(v_1,v_2)$
can in general be written:
\beq  v_1' = v_1-(v_1-v_2)(1-a)= av_1-(a-1)v_2 \label{vpostcol}\eeq 
Such an expression unifies elastic and dissipative collision through the dimensionless
parameter 
\beq a =\dfrac{m_t-\alpha m_b}{m_t + m_b} \label{defa}\eeq
$\alpha$ being  the restitution coefficient which describes the inelasticity of tracer-bath collisions (elastic if $\alpha =1$, and inelastic if $\alpha <1$ \cite{Brill}).  { It is remarkable that any system with dissipative collisions may therefore be mapped to an elastic system with a different mass ratio $m_t/m_b$, leading to the same dynamics and stationary states for the tracer \cite{Santos06,Pias06}. }
The ``memory-less'' situation $a=0$ is of particular interest due to its simplicity: 
each collision with a bath 
particle erases any memory of the tracer's previous state, and this case has 
been thoroughly studied and solved in different settings \cite{Pias83,Gervois,Santos06}. {The original Lorentz model corresponds to $a=-1$, or $a=-\alpha$ in its inelastic variant \cite{Martin07}, while the opposite Rayleigh limit $a=1$ (of an infinitely massive tracer) was considered in \cite{Eder88}.}

 The exponent $\nu$ introduced in (\ref{eq:nucoll}) encodes different scattering behaviors 
 \cite{Ernst06,Ernst81}; 
 The three most common models are $\nu=1$ corresponding to 
 Hard Rods, $\nu=0$ for Maxwell particles and $\nu=2$ for Very 
 Hard Rods, the latter two approaches being often invoked 
 to reproduce qualitatively some hard rod properties while 
 simplifying calculations \cite{Krap01}. 
 Their merits are discussed in \cite{Eder88} among others, but it will be shown 
 in later sections that the value of $\nu$ may affect significant properties 
 of the solution, some of which may exhibit crossovers depending on the value of $\nu$. {Negative values of $\nu$  have been shown to produce some interesting phenomenology in the Lorentz model, such as the remarkable runaway effect \cite{Pias81}, however our one-dimensional model with finite mass ratio produces well-behaved solutions if $\nu>-1$ and we shall restrict our study to this case.}
 
It should also be noted that the molecular chaos assumption underlying the Boltzmann
equation approach --motivated by the desire to derive analytical results-- prevents the occurrence of single-file diffusion \cite{Hahn}. However, when possible, we will discuss systems with a higher dimensionality, for which the molecular chaos assumption is justified in the low density 
limit.

It has been shown \cite{Alas10} that the velocity distribution quickly relaxes to a stationary solution 
$f_{\nu}=\lim_{t\to\infty} F_\nu$, on 
which we shall concentrate in the remainder: it obeys 
\beq g \dfrac{d f_{\nu}(v)}{d v} = \int_{-\infty}^{+\infty} dv_1 dv_2 \,  |v_1-v_2|^\nu \, f_{\nu}(v_1)\,\Phi(v_2)\left[\delta(v_1'-v) - \delta(v_1-v) \right] \label{bolstat}\eeq
which expresses the balance between the energy received from the external field $g$,
and the energy transfered to the bath through collisions. Eq. (\ref{bolstat}) may be rewritten as
 \beq g \dfrac{d f_{\nu}(v)}{d v} = r_\nu(v) \left[ H_\nu(v)-f_{\nu}(v)\right] \label{bolstat2}\eeq
 where 
\beq r_\nu(v) =\int_{-\infty}^{+\infty} du \,|u-v|^\nu \Phi(u) \label{defr}\eeq
The latter quantity is the velocity-dependent 
collision frequency in the gas, and $H_\nu(v)$ is given by
\beq r_\nu(v) H_\nu(v)= \int dv_1 dv_2\, |v_1 -v_2|^\nu \,f_{\nu}(v_1)\,\Phi(v_2) \delta(v'_1 -v) \label{defH} \eeq
which is the gain term of the collision operator, i.e. the transition rate toward velocity $v$ through the collision process, integrated over all possible initial velocities (dummy variable $v_1$).
Clearly, $f_{\nu}(v)=H_\nu(v)$ 
is the equilibrium solution for vanishing acceleration, when detailed balance entails that entering 
and leaving fluxes are equal at each point in phase space. In this field-free case ($g=0$), 
two situations discussed in \cite{Santos06} ensure that the tracer velocity distribution is directly given by the bath distribution: first, for memory-less collisions ($a=0$), we have
\beq f_{\nu}(v,0,0|\Phi)=\Phi(v)  \eeq
for arbitrary bath statistics, as can be checked from 
(\ref{defH}). Second, when the bath distribution is Gaussian 
$\Phi(v)=G(v) = (2\pi)^{-1/2} \exp(-v^2/2)$, one finds (see also \cite{Martin99})
\beq f_{\nu}(v,0,a|G) = G\left(\sqrt{\dfrac{1+a}{1-a} }v\right) \eeq
In general however, the tracer distribution differs from the bath distribution even in this unaccelerated limit, except in the tails (see section \ref{gen-bath}).

We will later need to consider the distribution of pre-collisional velocities, i.e. velocities sampled right before each collision rather than uniformly over time. This distribution is usually biased toward high velocities --which have enhanced collision rate as long as $\nu > 0$--  and it is given by \cite{Visco}
\beq  \tilde f_{\nu}(v) = \dfrac{r_\nu(v)}{\omega}f_{\nu}(v)\eeq 
where the mean collision frequency is defined as
\beq \omega=\int_{-\infty}^{+\infty} dv \,r_\nu(v)\,f_{\nu}(v) \label{defomega}\eeq

\subsection{Implicit form}
It proves convenient to recast the stationary probability density function $f_{\nu}(v)$ as an implicit
integral form. To this end, we write the velocity of the tracer at any given time as $v=v_1+gt$, 
with $v_1$ the velocity that was acquired during its last encounter with a scatterer, and $gt$ 
the contribution of the driving field during the timespan of ballistic flight $t$. We next consider 
the conditional probability that no collision occurs during $t$ for a tracer starting with velocity $v_1$, which involves the velocity-dependent collision frequency $r_\nu(v)$:
\beq P(t|v_1)=\prod_{k=1}^{t/\Delta\tau} \left[1-\Delta\tau\, r_\nu(v_0+gk \Delta\tau)\right] \too_{\Delta\tau \to 0}\exp\left(-\int_{0}^t d\tau \,r_\nu(v_1+g\tau) \right) \label{conditional}\eeq
We then need the probability density for the post-collisional velocity $v_1$.  By definition it is proportional to the gain term  $r_\nu(v_1) \,H_\nu(v_1)$ of the Boltzmann operator, which covers all collision events from which the tracer may emerge with velocity $v_1$ (proper normalization of the probability densities will be enforced {\it a posteriori}).
Finally, integrating over all possible time-spans $t\geq 0$, the stationary tracer velocity distribution 
can be written 
\beq \label{implicit} f_{\nu}(v)  =\int_{0}^{+\infty} dt  \, \exp\left(-\int_{0}^t d\tau \,r_\nu(v-g\tau) \right) r_\nu(v-gt) H_\nu(v-gt) \eeq
It can be checked by direct calculation that the above function obeys 
the Boltzmann-Lorentz equation \eqref{bolstat2}. A similar equation may be found for the time-dependent solution, with explicit dependence in the initial condition $F_\nu(v,0)$: 
\beq \label{implicitF} F_{\nu}(v,t)  = \int_{0}^{t} dt'  \, \exp\left(-\int_{0}^{t'} d\tau \,r_\nu(v-g\tau) \right) \left[r_\nu(v-gt') H_\nu(v-gt',t-t') + F_\nu(v-gt,0) \,\delta(t-t')\right] \eeq
where $H_\nu(v,t)$ is defined as in \eqref{defH} substituting $F_\nu(v,t)$ for $f_\nu(v)$. This expression is useful for studying the relaxation process toward the stationary solution $f_{\nu}(v)$ in the limit $t\to \infty$ \cite{Alas10}. The stationary solution may also be interpreted as an eigenfunction of a linear integral operator  $\mathcal L_\nu$, associated with the eigenvalue $1$:
\beq \label{operator} f_{\nu}(v)=\mathcal L_\nu[f_{\nu}](v) =\int_{-\infty}^{+\infty} du \,f_{\nu}(u)\,K_\nu(u,v)\eeq
where the kernel $K_\nu(u,v)$ is identified from equation \eqref{implicit}:
\beq \label{kernel} K_\nu(u,v) = \int_{0}^{+\infty} \!\!dt  \, \exp\left[ R_\nu(v-gt) -R_\nu(v)\right] \,\dfrac {|u - v + gt|^\nu} {(1-a)^{\nu+1}} \,\,\Phi\!\left(\dfrac{v-gt-au}{1-a}\right) \eeq
with 
\beq R_\nu(v) = \dfrac{1}{g}\int_{0}^v du \,r_\nu(u) \label{defR} \eeq
This form simplifies in the case of Maxwell particles ($\nu=0$)
\beq r_0(v)=\int du \,\Phi(u) = 1 \qquad\hbox{and}\qquad R_0(v) = \dfrac{v}{g} \eeq 
The kernel then reduces to a Laplace transform
\beq  K_0(u,v)=\dfrac{1}{1-a}\int_{0}^{+\infty} dt  \, e^{-t} \Phi\left(\dfrac{v-gt-au}{1-a}\right) \eeq 
If furthermore $a=0$, the solution itself becomes a transform of the bath distribution (as noted in \cite{Eder88,Alas10} in the special case of a Gaussian bath $\Phi(v)=G(v)$):
\beq f_{0}(v,g,0) = \int_{0}^{+\infty} dt  \, e^{-t} \Phi\left(v-gt\right)  \label{laplace}\eeq
For general values of $\nu$ and $a$ however, no obvious simplification can be found for the kernel $K_\nu(u,v)$. Expression \eqref{operator} nevertheless allows for efficient numerical solving, as we now describe.

\subsection{Numerical resolution of the Boltzmann-Lorentz equation}
\label{num-methods}

Three different techniques have been employed to obtain
numerically the tracer velocity statistics. We start by summarizing their 
main features, before testing their compatibility.

\subsubsection{Numerical integration and iterative solving}
\label{num-int}
The operator $\mathcal L_\nu$ derives from the Boltzmann-Lorentz equation, which by construction 
preserves the total probability in phase space. $\mathcal L_\nu$ must therefore conserve 
the integral of any function upon which it is applied. This implies that its spectrum is reduced to two eigenspaces, associated with eigenvalues $1$ (where the sought solution lies) and $0$ (with non-physical eigenfunctions having vanishing integral). The application of $\mathcal L_\nu$ therefore acts as a projection on the eigenspace associated with the eigenvalue $1$.

Numerically, discretizing $\mathcal L_\nu$ into a matrix $L_\nu$ and starting with a random vector, the iterative application of $L_\nu$ and subsequent normalization of the vector should converge toward the physical solution. Indeed all elements of $L_\nu$ are positive, therefore its largest eigenvalue is non-degenerate by the Perron-Frobenius theorem, associated with a positive eigenvector. This unique eigenvector coincides with the physical solution obtained through numerical simulations or determined analytically whenever possible, and this technique allows for fast computation of the solution to any precision, especially in the large velocity tails where other methods may become inefficient. 
At this point, we have to choose between two strategies: 
solving the eigenproblem numerically for the matrix $L_\nu$, or alternatively 
applying the matrix iteratively until proper convergence is achieved. 
We have checked that both routes 
are equally precise and efficient, for a similar computational cost.

\subsubsection{Gillespie algorithm}
The second numerical technique used relies on the Gillespie algorithm, adapted to the Boltzmann-Lorentz equation by Talbot and Viot \cite{Gillespie}. The salient features are as follows.
We recall from \eqref{conditional} the conditional probability that no collision occurs during 
time $t$ for a tracer starting with the post-collisional velocity $v_1$ 
\beq \ln \, P(t|v_1) = R_\nu(v_1+gt)-R_\nu(v_1)\eeq 
This relationship may be inverted (either analytically or numerically by dichotomy, the latter being necessary for $\nu \neq 0,2$ or with a non-Gaussian bath) to obtain $t(P(v_1+gt|v_1))$, allowing us to randomly generate collisions times for the tracer, after having drawn $P$ uniformly
in [0,1].

The velocity $v_2$ of the collision partner must then be generated with the correct weight $w(v')$. First, the flux of collisions is divided in two contributions, coming from the right and left of the tracer:
\beq r_+(v_1) = \int_{-\infty}^{v_1} du \,|v_1-u|^\nu \Phi(u)\eeq 
\beq r_-(v_1) = \int_{v_1}^{+\infty} du \,|v_1-u|^\nu \Phi(u)\eeq 
The probability of a collision coming from the right (resp. the left) is given by $r_+(v_1)/r(v_1)$ (resp. $r_-(v_1)/r(v_1)$). 
The direction of the next collision is thus determined using another random uniform number over $[0,1]$. Depending on which side is chosen, $w(v_2) = w_\pm(v_2)$ given by 
\beq w_\pm(v_2)=|v_2-v_1|^\nu \dfrac{\Phi(v_2)}{r_\pm(v_2)}\eeq 
The velocity $v'$ must be drawn with this weight, which is performed using an acceptance-rejection 
process: $v_2$ is generated according to the bath distribution $\Phi(v_2)$ (which significantly 
improves the efficiency in generating high velocities over a uniform generation with a cut-off), and then accepted with a probability 
\beq P_{acc}(v_2)=\dfrac{w_\pm(v_2)}{\Phi(v_2)}\dfrac{\Phi(v_0)}{w_\pm(v_0)}\eeq 
with $v_0$ the velocity for which $w(v_0)$ is maximal (which can be computed analytically for $\nu = 0,1,2$). 

\subsubsection{Direct Simulation Monte Carlo}
We also used Direct Simulation Monte Carlo (DSMC) \cite{Bird98} simulations as an alternative to the Gillespie algorithm, in order to experiment easily on various baths and collision laws, without 
performing the (bath-dependent) analytical computations which increase the efficiency of the Gillespie method. DSMC differs from the previous approach in that instead of generating the 
correct collision times directly, a collision partner is drawn at each 
step for the tracer,  with a velocity distributed according to the bath probability density function $\Phi(v_2)$. This collision event is then accepted with probability 
\beq P_{acc}=\dfrac{\Omega(v_2-v_1)}{\Omega_{max}} \eeq 
with $\Omega(v_2-v_1) =  |v_2-v_1|^\nu$ and $\Omega_{max}$ the maximal value encountered for $\Omega$ at this point. The time counter is incremented by $1/ \Omega(v_2-v_1)$ at each step. This procedure 
converges in a few hundred steps per particle toward a satisfactory evaluation of the correct collision frequency, as the observed velocity distribution for the tracer (over many realizations, and/or sampled uniformly over time) is then identical to the result of the Gillespie algorithm, 
or analytical computations whenever they can be performed.

 This method is not significantly less efficient than the Gillespie algorithm, despite its lesser
specificity. It can be extended to represent collisions in higher dimensionality
--as required for the discussion in section \ref{ssec:diff}-- 
by drawing a vector $\hat \sigma$ uniformly on the unit sphere and redefining
 \beq \Omega({\bf v_2}- {\bf v_1}) =  \sigma_1^{d-1} [(\bf v_2-\bf v_1).\hat\sigma ]^\nu \eeq 
with $\sigma_1^{d-1}$ the cross-section of the tracer-bath collisions. The post collisional velocity for the tracer is then given by
 \beq {\bf v'_1} = \mathbf{ v_1} + (1-a) \,[ (\bf v_2-\bf v_1).\hat\sigma ] \hat \sigma\eeq 

\subsubsection{Assessing the three numerical techniques}

\begin{figure}[htb]
\begin{center}
\includegraphics[width=0.7\linewidth, clip]{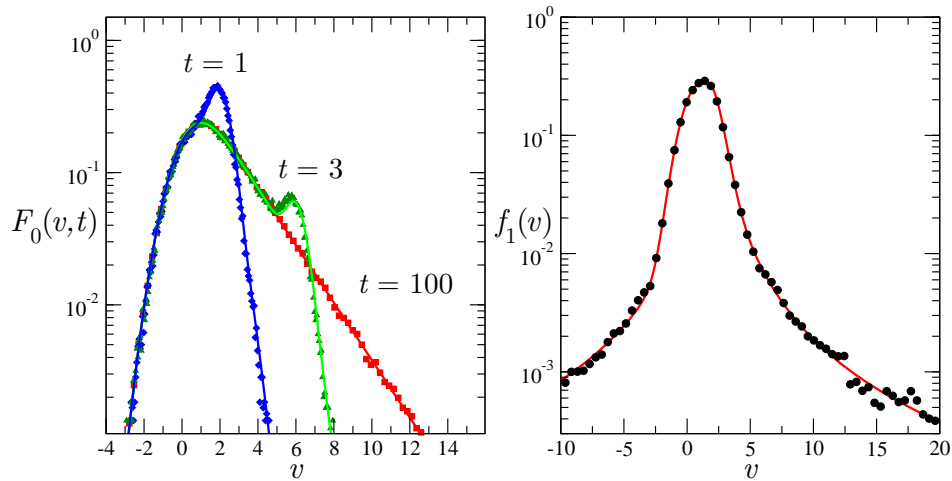}
\caption{Left: Time dependent velocity distribution $F_0(v,t,g=1,a=0)$, in a Gaussian bath $\Phi(v)=(2\pi)^{-1/2} \exp(-v^2/2)$ starting from a Gaussian initial condition with half variance.  Gillespie simulation results (symbols)
are compared to the analytic solution obtained in \cite{Eder88} (solid lines). Right:  stationary solution $f_1(v)$, with the same parameters and initial conditions, in a Lorentzian bath $\Phi(v)= \pi^{-1}/(1+v^2)$. The results from numerical integration (solid line) and DSMC simulation 
at large times
(symbols) are seen to be fully consistent.}
\label{simus}
\end{center}
\end{figure}

The implicit route allows for efficient computation of the stationary 
tracer velocity distribution with important precision.
Hereafter, it will be the method used whenever a specific prediction
requires unusual numerical precision (good sampling of the tails).
This will in particular be the case of Figs. \ref{Nonloc} and \ref{Nonloc2}.  
Among the three techniques, DSMC appears on the other hand to be
the more versatile, since a change in the model parameter 
like softness exponent $\nu$ does not lead to any complication. 
Finally, Gillespie and DSMC algorithms are well suited to
study the transient regime before the steady state is
reached (see e.g. \cite{Eder88,Soto,Alas10}), together with diffusive properties. 
Fig.~\ref{simus} demonstrates the validity 
and consistency of all three techniques in two settings.

\section{Velocity tails and diffusive behavior}
\label{gen-bath}

\subsection{Asymptotics}
\label{sec:asym}
As we shall disregard the time evolution of the system, the term "asymptotic behavior" refers here to the behavior of the stationary tracer distribution $f_{\nu}(v)$ for extreme values of the velocity.  Rewriting the implicit equation \eqref{implicit} as
\beq  f_{\nu}(v)  = \dfrac{1}{g}\int^{v}_{-\infty} du  \, \exp\left[R_\nu(u) -R_\nu(v)+ \ln H_\nu(u)\right]  r_\nu(u)  \label{asymplicit} \eeq
with $H_\nu$ and $R_\nu$ defined in \eqref{defH} and \eqref{defR}, {and assuming that both $\Phi$ and $f$ decay exponentially} we may use Laplace's method to approximate the integral over $t$. It is easily seen that
\beq 
R_{\nu}(v) \approx  \dfrac{|v|^{\nu+1} }{ (\nu+1) g}\qquad \text{when }v \to \infty
\eeq 
while we may choose a general stretched exponential expression for the tails of the bath distribution
\beq \ln \, \Phi(v) = -\dfrac 1 \mu \left| v \right|^\mu + O(\ln v)\label{firstmu}\eeq 
where $\mu > 0$, as found in various granular systems \cite{Montanero,Ernst02,Ernst06}. 
Power-law tails are obtained in the limit $\mu \to 0$, { while $\mu \to \infty$ corresponds to a uniform distribution with bounded support $[-1, 1]$. We shall restrict ourselves to $\nu>-1$ as $R_\nu$ becomes logarithmic when $\nu \to -1$ and we cannot expect an exponential decay of the integrand and apply Laplace's method anymore.}

We may gain intuitive insight into the large $v$ asymptotics by noting that competing
effects are at work: such velocities may be reached either through collision with sufficiently 
energetic bath particles, or through acceleration without collision for a sufficiently long time. 
The relative importance of these effects is thus determined on the one hand by the abundance of 
energetic particles in the bath, characterized by the parameter $\mu$, and on the other hand by 
the velocity-dependent collision frequency characterized by $\nu$.
Consequently, if $\nu$ is large enough compared to $\mu$, the increase in collision frequency 
for high velocities is so steep that ballistic flight is interrupted before the tracer can be 
significantly accelerated. The largest velocities are therefore reached through collisions with 
energetic scatterers, and the Boltzmann equation is dominated by the gain term, entailing that 
the tracer distribution is pushed back toward the bath distribution. This may be seen 
as thermalization with the bath tails. If on the other hand $\mu$ is sufficiently large
compared to $\nu$, the bath tails are relatively depleted, while ballistic flight and acceleration 
are less impeded by the collision process. This corresponds to a prevalence of the loss term 
in the master equation: it is more probable for the tracer to have a large velocity before 
a collision (due to acceleration) than after it.

A more refined asymptotic analysis confirms the above qualitative features.
Since the argument of the exponential in \eqref{asymplicit} decreases as $u \to -\infty$, either
the global maximum is located on the boundary at $u=v$, or it corresponds to some local extremum $u=u_0$ such that the derivative of the argument vanishes
 \beq \dfrac{1}{g}\, r_\nu(u_0)  +  \dfrac{d\ln H_\nu}{d u}(u_0) = 0 \eeq  

{
If  $u =u_0$ the $v$ dependence of the integral receives no contribution from the $\ln H_\nu (u)$ term --the gain term of the Boltzmann equation is negated-- and we have $f_\nu(v) \sim \exp[-R_\nu(v)] $. This is possible only in the positive tail : when $v\to-\infty$, all the local extrema eventually exit the interval $[-\infty,v]$ and the global maximum is necessarily located at the boundary $u=v$, regardless of the parameters $\mu$ and $\nu$.

In the latter situation (for either tail), $R_\nu$ disappears from \eqref{asymplicit}  --the loss term is negated-- and we have
\beq  f_{\nu}(v) \sim H_\nu(v) =\dfrac{1}{r_\nu(v)(1-a)^\nu} \int_{-\infty}^{+\infty} du\, |u -v|^\nu \,f_{\nu}(u)\,\Phi\left(\dfrac{v-au}{1-a}\right) \label{defH2}\eeq
We search for self-consistent solutions using Laplace's method, and find
\beq f_{\nu}(v)  \sim \Phi(\lambda v)\label{eq:lefttail}\eeq
as in the unaccelerated problem, where the dilation coefficient $\lambda$ satisfies
\beq 
\lambda=\dfrac{\left| 1-|a|^{\mu/(\mu-1)} \right|^{(\mu-1)/\mu}}{1-a} \quad \text{ and } \quad \lambda=\dfrac{a}{1-a} \quad \text{ if } \quad \mu=1
\eeq 
For a Gaussian bath (for which  $\mu=2$) it becomes $\lambda^2 = (1+a)/(1-a)$ which equals 
$m_t / m_b$ when collisions are elastic, as expected from thermalization. 

The other solution to \eqref{defH2} is $f_\nu(v)  \sim f_\nu(v/a)$, provided that $\Phi(v)$ vanishes faster than $f_\nu(v/a)$ so that the stationary point is asymptotically close to $v=au$. This solution is self-consistent only if $a<0$, as it then relates one tail to the other rather than equating the same tail in two different points.

Combining these observations, it appears that the leading behavior for $v\to + \infty$ is simply controlled by the value of the integrand of \eqref{asymplicit} at each of the two locations $u=u_0$ and $u=v$, and switches from field-driven (due to the loss term) if the first dominates to thermalized if it is the second (see Table~\ref{tab}). For $v\to -\infty$, however, the default behavior is thermalized, unless $a<0$ and the positive tail is field-driven. In that case, collisions may revert large velocities, and the left tail becomes the image of the right tail with $v\to av$.}

\begin{table}[h]
\begin{center}
\halign{\hspace*{60pt}\hfil  # & # \hfil \cr
\begin{tabular}{clrc}
\noalign{\hrule height 0.8pt }\noalign{\smallskip}
 $\mu < \nu +1$ & \multicolumn{2}{c}{$\mu = \nu +1$} & \hspace*{-10pt}$\mu > \nu +1 $ \vspace*{2pt} \\
 & $g<\lambda ^{-\mu}$ & $g>\lambda ^{-\mu}$  \\
$\downarrow$ &$\swarrow $&$ \searrow$ & $\downarrow$\\
\hline \noalign{\smallskip}  \multicolumn{2}{l}{

\begin{tabular}{cl} $f_{\nu}(v)$& \hspace*{-10pt}$\sim \Phi(\lambda v)$ \vspace*{2pt}\\
&\hspace*{-10pt}$\sim e^{-|\lambda v|^\mu/\mu}$
\end{tabular}

} & 

\multicolumn{2}{r}{\begin{tabular}{rl} $f_{\nu}(v)$& \hspace*{-10pt}$\sim e^{-R_\nu(v)}$ \\
&\hspace*{-10pt}$\sim e^{- v^{\nu+1}/(\nu+1)g}$
\end{tabular}}

\vspace*{5pt}\\
\noalign{\hrule height 0.8pt}
\end{tabular}
&\hspace*{25pt}
\parbox[c]{1em}{\includegraphics[width=18\linewidth, clip]{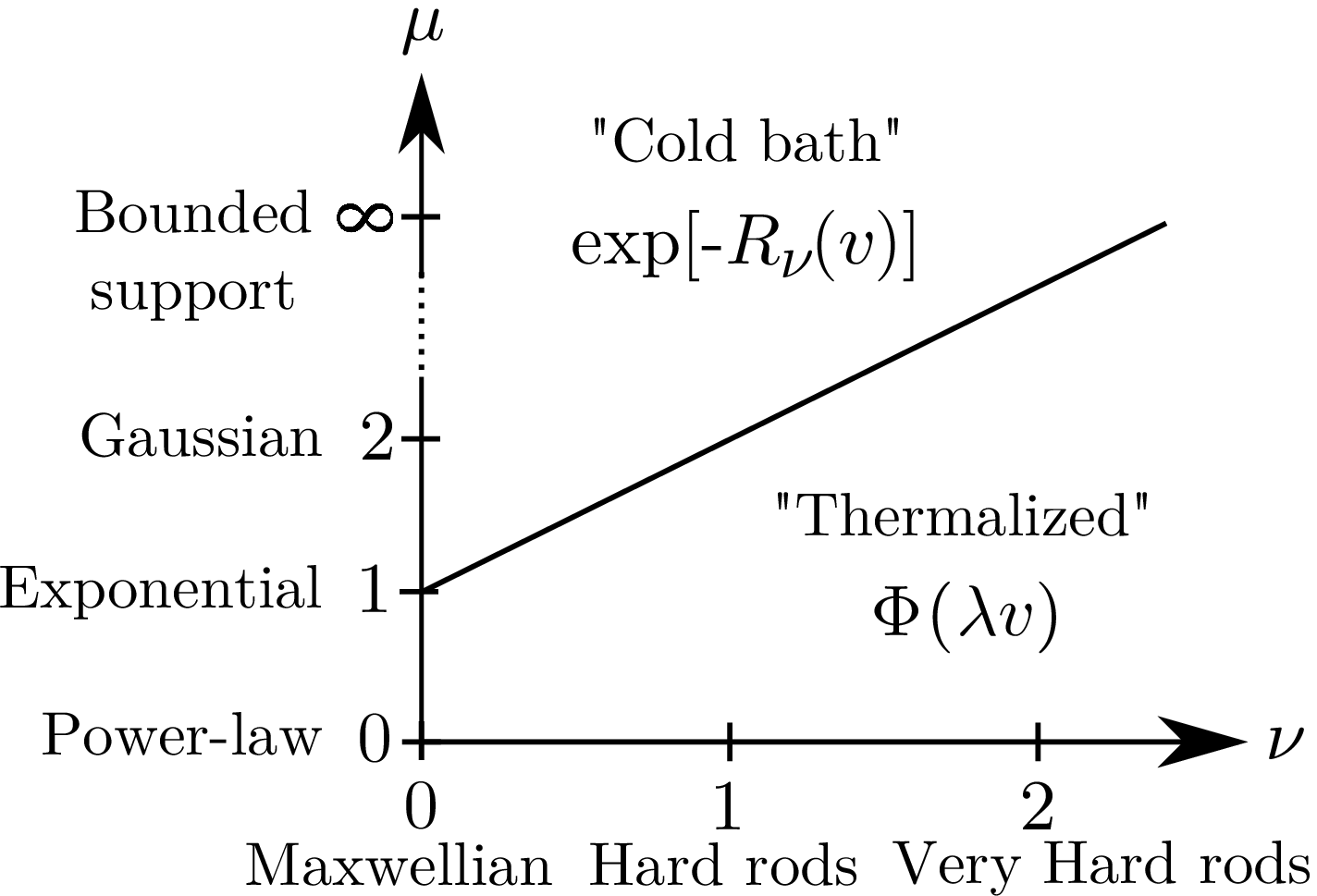}}\cr}
\vspace*{3pt}
\end{center}

\caption{Asymptotic behaviors for the positive tail $v\to+\infty$: thermalized (left column of the table) or field-driven (right column). The negative tail is either its image with $v\to av$ in the field-driven case with $a<0$, or thermalized in any other case. {This analysis can be extended to any $\nu > -1$, including the three usual values $\nu=0,1,2$ as shown on the right-hand graph.}}
\label{tab}

\end{table}

\begin{figure}[htb]
\begin{center}
\includegraphics[width=0.9\linewidth, clip]{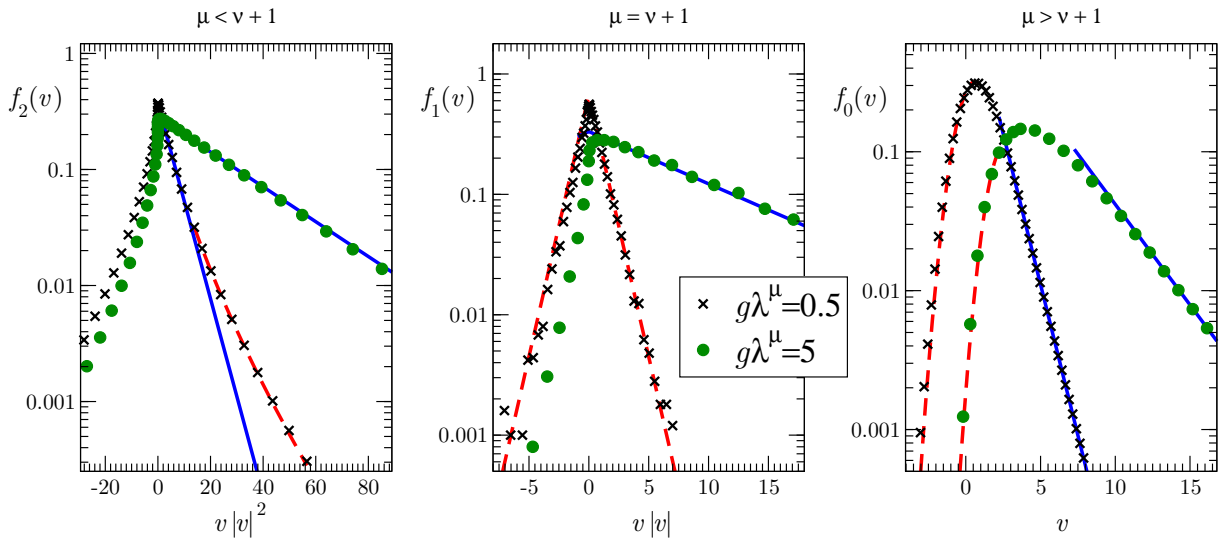}
\caption{Asymptotic behavior in a Gaussian bath ($\mu=2$). From left to right: $f_{2}(v)$, $f_{1}(v)$ and  $f_{0}(v)$, computed from DSMC simulations for $a=-1/3$, $g=1$ (crosses) and $a=1/3$, $g=3$ (dots). The solid line is the cold bath behavior $f_{\nu}(v) \propto \exp\left[ -v|v|^{\nu}/(\nu+1)g \right]$, while the dashed line is the "equilibrium" asymptotic $f_{\nu}(v) \propto  \Phi(\lambda  v)$. For $f_{2}(v)$ (left), this asymptotic expression describes the positive tail, although the value of the threshold $v_c = g\lambda^\mu$ may be too high for the crossover to be easily observable (as seen here with the dots which retain the cold bath behavior for the 
whole  velocity range reported). { Conversely, the negative tail is thermalized by default, and appears so even in the lone case $a<0$, $\nu+1 < \mu$ where it must become cold bath-like. This asymptote is eventually reached for lower velocities, as evidenced in Fig.\ref{maxnul}.} }
\label{vstar}
\end{center}
\end{figure}

The large velocity scenario sketched in Table \ref{tab} is as follows.
If $\nu +1 > \mu$, we recover the bath behavior in $\exp(-v^\mu)$,
in agreement with the intuitive argument outlined above, while if 
$\nu +1 < \mu$, the positive velocity tail exhibits a $\exp(-v^{\nu+1})$ decay,
hereafter referred to as the cold bath behavior. These expectations, {disregarding subleading terms such as polynomial prefactors},
are fully corroborated by Fig. \ref{vstar}.

If $\nu+1 =\mu $, both behaviors correspond to the same exponent, and prevalence of one process 
over the other depends on the value of the driving field $g$ compared to $ \lambda ^{-\mu}$. 
For a Gaussian bath with $\mu=2$, the borderline case corresponds to the hard-rod model, as depicted in the middle 
picture of Fig. \ref{vstar}: upon increasing the acceleration, the positive tail switches 
continuously from the field-independent expression $ \exp\left[-(\lambda v)^2/2\right]$ 
when $g\lambda^2 <1$ to the field-dependent $ \exp\left[-v^2/2g\right]$ when $g\lambda^2 >1$.
The critical field, $1/\lambda^2$, reduces to unity in the memoryless case $a=0$.

{ If $\nu +1 > \mu$, the onset of the positive tail behavior described above is given by the point where \beq v^{\nu +1} \sim g \lambda^{\mu} v^\mu \quad  \Rightarrow  \quad v \sim v_c = \left( g \lambda^\mu \right)^{(\nu +1- \mu)^{-1}}  \label{threshold}\eeq  Assuming this threshold $v_c$ is larger than the typical bath velocity $v_b$ (set to unity), there is an intermediate asymptotic range $1\ll~v\ll~v_c$ dominated by acceleration. The bath-like tail is thus expected to appear only when $v \gg v_c$. If however $\nu+1 < \mu$,  both the intermediate range and the (positive) tail are similarly field-driven, and no such clear-cut crossover can be observed. 

Both cases are represented in Fig. \ref{vstar} : the latter appears on the right-hand side, and the former on the left-hand side, where the crossover from the field-driven behavior to the bath-driven tail may be seen on the curve with $v_c=g \lambda^2 = 0.5$ (crosses). The intermediate asymptotics may nevertheless dominate over a large range of positive velocities, as can be seen for a higher value of $v_c =5$ (filled dots) where the thermalized tail has yet to appear.

In the strong-field limit $g\to \infty$, the threshold $v_c$ diverges and the field-driven behavior prevails for all $v\gg 1$ over the whole parameter space of $\mu$ and $\nu$. As can be seen from our choice of dimensionless variables in section~\ref{sub:stat}, this is equivalent to taking the limit of motionless bath particles $v_b \to 0$ while keeping $v$ constant : this limit $\Phi(v) \to \delta(v)$ is the so-called "cold bath" that will be discussed in section \ref{Coldbath} and exhibits characteristic field-driven tails.}

The opposite limit $g\to 0$ is singular as it suppresses this cold bath (field-driven) behavior altogether, 
rendering our discussion of limiting cases invalid as the only possible asymptotic behavior is 
the thermalized expression $f_\nu(v) \sim \Phi(\lambda v)$ in both tails.

\subsection{Diffusive properties}
\label{ssec:diff}
Another point of interest lies in the tracer diffusion coefficient $D$, which was computed 
in \cite{Alas10} for a Gaussian bath and $a=0$, using both its 
Green-Kubo expression and its definition in the hydrodynamic diffusion mode of the inhomogeneous 
Boltzmann-Lorentz equation. These two methods were shown to give identical results for $\nu = 0$ and $\nu=2$. The second method was also applied in \cite{Soto} to the case $\nu=1$ and $a=0$ with a dichotomous bath, revealing the complex dependence of the diffusion coefficient on the magnitude of the field, with a minimum at finite $g$, while this coefficient $D$ was shown to be monotonically increasing for $\nu=0$ and decreasing for $\nu=2$ in the aforementioned discussion of the Gaussian bath \cite{Alas10}.
Here we consider an arbitrary bath distribution and parameter $a$, and observe once again 
that the effects of acceleration and thermalization disentangle when $\nu=0$. We also demonstrate 
that these results are qualitatively unchanged in higher dimensional space, as reported for other 
observables in \cite{Martin99}.

\begin{figure}[htb]\begin{center}
\includegraphics[height = 150 pt, clip]{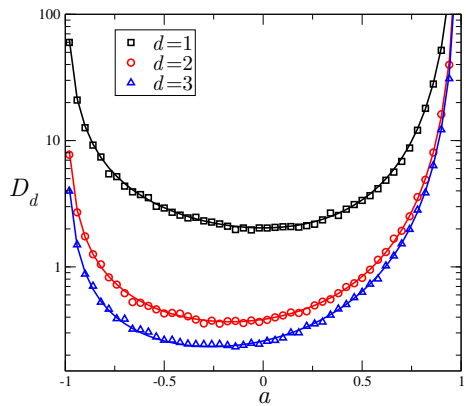}
\caption{The diffusion coefficient $D_d$ for $d=1,2,3$ and $g=1$ for Maxwell particles in a Gaussian bath, against the parameter $a$. The symbols represent DSMC simulations, while the solid lines are obtained using equation \eqref{diffd}.}
\label{diffusion}\end{center}
\end{figure}

As shown in the appendix, letting $\nu=0$ allows us to relate the diffusion coefficient to the variance of the tracer velocity through a simple combination
\beq 
D=\dfrac{\left\langle v^2 \right\rangle - \left\langle v \right\rangle^2}{1-a}  
\eeq 
This variance may easily be determined recursively from that of the bath, by integration over the Boltzmann-Lorentz equation
\beq \langle v^{n}\rangle-g\,n\,\langle v^{n-1}\rangle =\lambda^{-n} \int_{-\infty}^{+\infty} du \, f_{0}(u)\, \phi_n(u) \eeq 
with $ \phi_n(u) =  \dfrac{1}{a-1} \int_{-\infty}^{+\infty} dv \, v^n \, \Phi \!\left(\dfrac{v-au}{a-1}\right)$ the $n$th moment of the displaced bath distribution.
 Therefore, if the bath velocity distribution is centered and has finite variance,
\begin{equation} D=\dfrac{1}{1+a}\left(\left\langle v^2 \right\rangle_b + \dfrac{g^2}{(1-a)^2}\right)\label{Deq}\end{equation}
where $\left\langle v^2 \right\rangle_b$ is the mean velocity squared in the bath, chosen equal to $1$ in our dimensionless variables. This result is in agreement with \cite{Alas10} in the limit $a \to 0$. It may be generalized to $d$-dimensional space (see Appendix A)
\begin{equation}D_d = \dfrac{d}{A_d(1+a)}\left(1 + \dfrac{g^2 d^2}{A_d^2(1-a)^2}\right) \label{diffd}\end{equation}
where $A_d=2 \pi^{d/2}/\Gamma(d/2)$ is the surface area of a $d$-dimensional hypersphere. 
This prediction is in very good agreement with the DSMC simulation data shown in Fig. \ref{diffusion}. 
In order to study the diffusive 
behavior of the tracer in these otherwise homogeneous simulations, its spatial coordinate 
is computed as the time-integral of the velocity. Between two collisions, the "space counter" in the direction of the acceleration is thus incremented by $vt + \frac{1}{2}g t^2$ with $v$ the velocity of the tracer during this free flight and $t$ the previously determined collision time.

\section{The cold bath}
\label{Coldbath}

In our previous discussion of the velocity tails for the tracer in a general bath, we have observed the prevalence of a "cold bath" asymptotic behavior for certain values of the parameters, see Table \ref{tab}.
The cold bath refers to a particular limit of the general Lorentz problem, where the scatterers are
motionless and their velocity distribution is represented by the Dirac delta function: it may be
understood as the limit $v_b \to 0$ as discussed in \ref{sec:asym}.

This setting has been
studied for hard rods $\nu=1$ with $a = 0$  \cite{Pias83} and in some limiting cases with $a \leq 0$
\cite{Martin99}. Inserting $\Phi(v) = \delta(v)$ in \eqref{defr} gives $r_\nu(v)=|v|^\nu$ and the
Boltzmann-Lorentz equation takes the form of a functional-differential equation
\beq gf'_{\nu}(v,g,a)=|v|^\nu \left(\dfrac{1}{|a|^{\nu+1}}f_{\nu}\!\left(\dfrac{v}{a},g,a\right)-f_{\nu}(v,g,a) \right) \label{cold}  \eeq
which is reminiscent of relations found in other stochastic processes such as random walks and growth models on trees \cite{Dean05}. In the field-free case, this equation reverts to a class of random collision problems systematically studied by ben-Avraham et al. \cite{Ben03}.


This system plays a prototypical role in the study of the general Lorentz problem for two reasons. { First, it may be seen as the strong field limit ($g\to\infty$) of any bath distribution as discussed in section \ref{sec:asym}. Second, there are two exact mappings allowing to reduce a more general problem to the exactly solvable case of a Maxwellian tracer in a cold bath, with both $\nu=0$ and $\Phi(v) = \delta(v)$, if only either one of these conditions is verified.} Indeed, given a cold bath distribution, we may derive the solution for any value of the collision kernel exponent from the Maxwellian solution $\nu=0$. This may be seen if we define
\beq  f_{\nu}(v)= h_0\left(\dfrac{v |v|^\nu}{\nu+1} \right)\eeq 
which verifies
\beq g  h'_0(v)=\dfrac{1}{|a|^{\nu+1}}  h_0\!\left(\dfrac{v}{ a |a|^\nu}\right)-  h_0(v) \eeq 
With the simple rescaling $a|a|^\nu\to a $ this equation is made to coincide with \eqref{cold}, 
which is associated to Maxwell particles. 

 On the other hand, for Maxwell particles ($\nu = 0$) a theorem proved by Wannier \cite{Wannier} for the three-dimensional case and Eder and Posch \cite{Eder88} in one dimension gives the complete (time-dependent) solution for any bath distribution $\Phi$ as a convolution over the cold bath solution
\beq F_{0}(v,t,g,a|\Phi)= \int_{-\infty}^{+\infty} du\, \Phi(u) \,F_{0}(v-u,t,g,a |\delta)\eeq 
This relation holds as well for the stationary state $f_{0}(v)$ which is under scrutiny here.

 However Wannier's convolution theorem does not hold for arbitrary values of $\nu$, which prevents from extending this mapping to settings where both $\nu$ and $\Phi$ are chosen different from the prototypical case ($\nu=0$, $\Phi(v)=\delta(v)$). Yet, this model still preserves many significant 
features of the general solution, as discussed previously.

\subsection{Explicit solution} 
\label{Colds1}

The cold bath equation \eqref{cold} may be rewritten as the implicit equation
\beq 
f_{0}(v)  =\frac{1}{g} \int_{-\infty}^{+\infty} du \,  f_{0}(u) \, \exp\left[- \dfrac{v-au}{g }\right] \, \Theta(v-au) 
\label{coldim}
\eeq
where $\Theta$ denotes the Heaviside function. This expression allows us to derive straightforwardly the solution for the "memory-less" case $a=0$ 
\beq f_{0}(v,g,0)  =\dfrac{ 1 }{g} \,\Theta(v)\,\exp\left[- \dfrac{v}{ g} \right] \label{colda0}\eeq
where it appears that the tracer velocity distribution has support on $v>0$. {The solution for any $\nu\neq 0$ may be recovered from the aforementioned mapping, in order to study the $\nu$ dependence of quantities of interest, such as the mean velocity (note the choice of index $\nu-1$ to simplify this expression)}
\beq  \langle v \rangle_{\nu-1} (g, a=0)  = \dfrac{1} {2\sqrt{\pi}} \,\Gamma\left(\dfrac{2+\nu}{2\nu}\right)\, (4\nu g)^{1/\nu}\eeq 
where $\Gamma$ is the Euler gamma function. Thus $\langle v \rangle_\nu \propto g^{1/(\nu+1)}$ in agreement with the dimensional argument presented in \cite{Pias83}. It should be noted that, unless $\nu=0$, this implies a breakdown of linear response: for $\nu=1$, $\langle v \rangle \propto g^{1/2}$ as discussed in \cite{Martin99}.

We now derive the explicit solution for $a\neq 0$, considering Maxwellian particles (without loss of generality, as argued above) which verify
\begin{equation}
gf'_0(v)= \dfrac{1}{|a|}f_{0}\!\left(\dfrac{v}{a}\right)-f_{0}(v) 
\label{coldMax}
\end{equation}
The general expression of the solution may be found by taking the Fourier transform of the function
\beq \hat f_{0}(s)= \int_{-\infty}^{+\infty} f_{0}(v) e^{isv} dv\eeq 
which in turn verifies
\beq igs\,\hat f_{0}(s)= \hat f_{0}(sa)-\hat f_{0}(s)\eeq 
\beq \hat f_{0}(s) = \dfrac{\hat f_{0}(sa)}{1+igs}\eeq 
Then by recurrence { we may relate $\hat f_0(s)$ to $\hat f_0(sa^{n+1})$. As $|a|<1$, $sa^{n+1} \to 0$ when $n\to\infty$. Furthermore $\hat f_{0}(0)=1$ due to the normalization of $f_{0}(v)$, hence}
\beq \hat f_{0}\left(s\right) =\lim_{n\to\infty}\dfrac{\hat f_{0}\left(s a^{n+1}\right)}{ \prod_{k=0}^n  \left(1+igs a^{k}\right)} = \prod_{k=0}^\infty  \left(1+igs a^{k}\right)^{-1} \label{coldhat}\eeq
Taking the inverse transform (with some care as the position of the poles and the integration contour depend on the sign of both $a$ and $v$)
\begin{align}&  f_{0}(v) =\sum_{k=0}^{\infty} \dfrac{ e^{-v/ga^k }}{ga^k}  \Theta(v/ga^k) \prod_{j \geq 0, j\neq k} (1-  a^{j-k})^{-1} \label{coldsol}\end{align}

\begin{figure}[htb]
\begin{center}
\includegraphics[width=0.4\linewidth, clip]{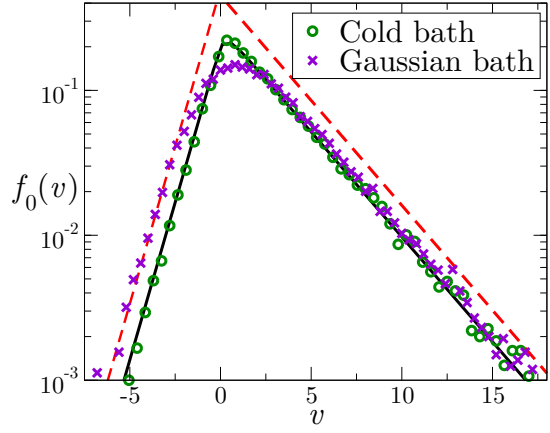}
\caption{Asymptotic behavior for a Maxwellian tracer in a cold bath and a Gaussian bath 
($a=-1/3$, $g=3$). The symbols are Gillespie simulation results, the solid line is the exact 
cold bath solution \eqref{coldsol}, and the dashed lines represent the asymptotic slopes $\exp(-v/g)$ (right, see Eq. \eqref{cold+}) and $\exp(-v/ag)$ (left, see Eq. \eqref{cold-}). {
Here, as $a<0$ and the positive tail is field-driven, the negative tail for the Gaussian bath is also similar to the cold bath solution, although that solution would vanish for negative velocities if $a\geq 0$ and the left tail would then be thermalized (see Fig. \ref{vstar}).} }
\label{maxnul}
\end{center}
\end{figure}

From this expression, we may identify the asymptotic behavior for $|v|\to \infty$ as the first non-zero term in the sum depending on the sign of $v$ and $a$.  We therefore obtain the characteristic "cold bath" positive tail behavior presented in Table \ref{tab}: when $v\to +\infty$,
\beq 
f_0(v)  \propto \exp\left[-\dfrac{v}{g}\right] \label{cold+}
\eeq
The negative tail behavior however differs from the situations discussed in section \ref{gen-bath}, 
where it was determined by the bath distribution. It is readily seen from \eqref{coldsol} that negative velocities cannot be reached if $a>0$ : in that 
case, equation \eqref{vpostcol} entails that the tracer velocity is reduced but never reversed 
by collisions with bath particles. Once it has become positive due to acceleration, it may not 
become negative again, which ensures that the stationary solution has support on $\mathbb R^+$ only. If however $a<0$, the negative tail is given by 
\beq 
f_0(v\to -\infty,g,a<0)  \propto \exp\left[-\dfrac{v}{a g}\right]  
\label{cold-}
\eeq
These asymptotic expressions for $|v|\to\infty$ with $a<0$ are corroborated by numerical results in 
Fig. \ref{maxnul}.

The moment hierarchy is obtained by a straightforward integration, and its general term is found 
\beq \langle v^n\rangle = n!\, g^n b_n \label{coldmoments}\eeq
where $b_n=1/(a,a)_n$ using the $q$-Pochhammer notation \cite{Gasper}
\beq (a,q)_n=\left\lbrace \begin{array}{lr}
1&n=0\\
\prod_{m=0}^{n-1} (1- a q^{m})&n\ge 1
\end{array}
\right. \label{defbn}\eeq 
This symbol appears in combinatorics as a generalization of the Pochhammer symbol or rising factorial: \beq \lim_{q\to 1^{-}}\dfrac{(q^x,q)_n }{(1-q)^n} = x(x+1)...(x+n-1) = \dfrac{\Gamma(x+n)}{\Gamma(x)}\eeq 
The expression $(a,a)_n$ converges in the limit $n\to \infty$ to a standard function of $a$ known as the Euler function (not to be confused with the better known Euler totient function in number theory).

 We may rewrite the cold bath solution \eqref{coldsol}, using the $b_n$ notation for the product over $j$, and recognizing the exponential term as the solution for $a=0$ in \eqref{colda0} where the acceleration has undergone the rescaling $g\to ga^k$. We obtain the concise formula
\beq f_{0}(v,g,a)=b_\infty   \sum_{k=0}^\infty  (-1)^k\, a^{k(k+1)/2}\,b_k \,f_{0}(v,ga^k,0) \label{coldseries}\eeq
This structure is interesting in several respects: first, the sum converges very rapidly for 
the needs of numerical computation, and is found in excellent agreement with our simulation 
results. Second, due to the simple form of $f_{0}(v,g,0)$, using Wannier's theorem to derive 
the full solution for Maxwell particles in any bath will only involve taking Laplace transforms 
of the bath distribution. An application to the Gaussian bath is given below. Finally, even though 
this structure will not be preserved for systems with arbitrary bath distribution and $\nu \neq 0$, 
it suggests that some significant traits of the general solution may be contained in the limiting 
case $a=0$, which is more readily solved in any model, see e.g. \cite{Gervois,Alas10}. However, 
both analytical expressions \eqref{coldsol} and \eqref{coldseries} shed little light on the behavior of the cold bath solution for 
positive $a$ in the limit $v\to 0$, which will warrant a separate investigation in 
section \ref{non-negative}.

\subsection{Application to the Gaussian bath}
We now revisit some properties of one of the most extensively studied 
settings, the Maxwellian tracer in a Gaussian bath ($\nu=0$, $\mu=2$) \cite{Gervois,Eder88,Alas10}. This will serve to 
demonstrate the application of results from the cold bath limit to arbitrary bath distributions, 
which we recall is exact for $\nu=0$. 

From Wannier's convolution theorem we deduce that the solution $f_{0}(v,g,0)$ in this setting is simply given by the Laplace transform of the Gaussian $G(v)$, as we previously saw in \eqref{laplace} (see also \cite{Alas10})
\beq 
f_{0}(v,g,0|G) = \dfrac{1}{2g}\,\exp\left[\dfrac{1}{2g^2}-\dfrac{v}{g} \right]\,\left(1+\erf\left[\dfrac{v-g^{-1}}{\sqrt 2} \right]\right)  \label{maxa0}
\eeq
Inserting this expression in the series \eqref{coldseries} allows for accurate numerical computation as well as asymptotic analysis for $a\neq 0$. 

We thus recover a result that was obtained in Ref. \cite{Eder88}, where use was made of a method 
specific to this choice of bath and collision kernel: using Mehler's formula \cite{Mehler}, 
it is possible to compute the analytical solution of the Boltzmann-Lorentz equation as 
an expansion in terms of the Hermite Polynomials $H_n(v)$
\beq f_{0}(v|G) =G(\lambda v) \, \sum_{n=0}^{\infty} \left(\dfrac{ \lambda g}{\sqrt 2}\right)^{\! n} \, b_n \, H_n \! \left(\frac{\lambda v}{\sqrt 2}\right) \label{maxstat}\eeq
However, this expansion diverges for any value of the parameters, even at vanishing velocity as $H_n(0) = (-1)^{n/2}\, n!/(n/2)!$ for even $n$, while $H_n(0)=0$ for odd $n$. Consequently, the expansion must be put under a different form to allow for numerical computation. It is suggested in \cite{Eder88}
that, owing to a theorem by Euler
\beq b_n = b_\infty \sum_{k=0}^\infty (-1)^k\, a^{nk}\, a^{k(k+1)/2} \,b_k\eeq 
from which, interchanging the sums, one finds as expected
\beq  f_{0}(v,g,a) =b_\infty  \sum_{k=0}^\infty (-1)^k\  a^{k(k+1)/2}\, b_k \,f_{0}(v,ga^k,0) \eeq 

Finally, we may be interested in seeing how exactly the hierarchy of moments with a Gaussian bath relates to its cold bath equivalent. We compute the moment-generatrix, using the connection between the Hermite polynomials and the derivatives of the Gaussian
\beq M(t)=\int_{-\infty}^{+\infty} dv \,e^{tv} f_{0}(v)= \exp\left[\dfrac{ t^2}{2\lambda^2}\right] \sum_{j=0}^k \dbinom{k}{j}	 \,\zeta^{(k-j)}(t)\,\left(i\sqrt 2 \lambda  \right)^{-j}\,H_j\!\left(\dfrac{i\,t}{\sqrt 2 \lambda}\right) \eeq 
with \beq \zeta(t)=\sum_{n=0}^{\infty} t^n \, g^n \,b_n\eeq 
from which we derive a general expression for the $k$-th moment of $f_{0}(v)$:
\beqa \langle v^k \rangle &=& \dfrac{d^kM}{dt^k} (t=0) \nonumber \\   &=& \sum_{j=0}^{\lfloor k/2 \rfloor}\, \dfrac{k!}{j!} \left(2 \lambda^2   \right)^{-j} \, g^{k-2j}\, b_{k-2j} 
\eeqa
The term $j=0$ (which is the leading order for large $g$) is recognized as the corresponding moment in the cold bath \eqref{coldmoments}, while the following terms appear to be specific to the Gaussian setting.

\section{Non-negative stationary distribution}
\label{non-negative}
The solution for $a>0$ exhibits some peculiar properties in the cold bath configuration: 
the tracer velocity can be reduced but never reversed by a collision. If the tracer reaches a positive velocity at a given time due to the positive acceleration, it will subsequently 
never be able to reverse its direction of motion. As mentioned in section \ref{Colds1}, this means
that the stationary distribution has support on the positive 
semi-axis only. This gives a new limit to consider, that of small velocities $v\to 0$, which has no 
equivalent in any of the settings previously considered, and may appear only for a bath distribution with support on an interval bounded from below.

\subsection{Heuristic argument}
We shall now show that the asymptotic behavior of $f_{0}(v)$ for vanishing velocities is 
of log-normal form. We start with a heuristic argument.
In order to reach a very low velocity $v$, numerous collisions must happen in a very short time 
span, before the acceleration can restore the velocity to its typical scale. Each collision multiplies the velocity of the tracer by $a$, therefore a sequence of repeated collisions may be viewed as a multiplicative random walk, or an additive random walk on $\ln v$:
$\ln v_{i+1} \approx \ln v_i + \ln a
$
where $v_i$ is the velocity of the tracer following collision $i$. After a random but large number of steps (collisions occurring almost instantaneously), we expect $\ln v$ to be normally distributed.

\begin{figure}[h]
\begin{center}
\includegraphics[height=160 pt, clip]{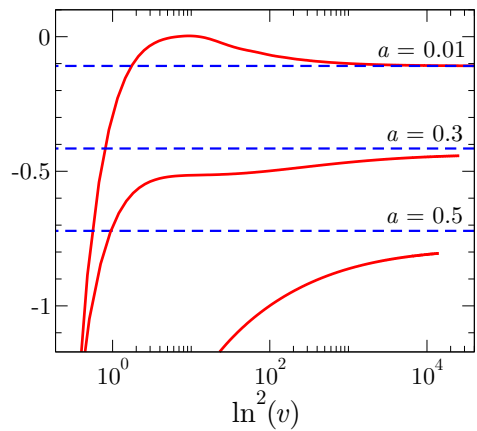}
\caption{Plot of $\ln f_{0}(v) / \ln^2(v)$ following from \eqref{coldsol}, as a function of $\ln^2(v)$ (increasing as $v\to 0$), for different values of $a$ and $g=1$. To each set is associated a horizontal line marking the expected limit $\chi/2=(2\ln a)^{-1}$, see Eq. (\ref{heurist}), from which we may see the approach toward the limit becoming extremely slow as $a$ increases.}
\label{Nonloc}
\end{center}
\end{figure}

This intuition may be put on more solid grounds. In any sequence that allows the tracer to reach a very low velocity, collision $i+1$ must take place in a time interval $\Delta t_i$ short enough that $g \Delta t_i \ll v_{i-1}$, lest the tracer be reaccelerated to its former velocity. This interval, the upper bound for the time span between two successive collisions, becomes smaller as lower velocities are involved, therefore we may choose a bound of the form $g \Delta t_i = \epsilon v_i$ with a constant factor $\epsilon \ll 1$.
At least one collision must occur during each interval $\Delta t_i$, which
has probability $1-\exp(-\omega \Delta t_i) \approx \omega \Delta t_i$. Starting from a characteristic velocity $\bar v$, we need $n$ such collisions to reach the velocity $v_n= a^n \bar v$, and we may associate to this sequence the following probability density
\beq 
f_{0}(v_n)  \propto \prod_{i=0}^n \omega \Delta t_i \approx \left( \dfrac{\epsilon \omega}{g} \right)^n\, a^{n(n+1)/2}\eeq 
Writing $\chi = 1/\ln \,a$, we have
\beq 
f_{0}\!\left(v\right) \sim \exp\left\{ \dfrac{\chi}{2}   \ln^2(v) \right\} \label{heurist}
\eeq
We cannot expect to obtain more than the leading order in $\ln^2(v)$ from such simple considerations, but Fig. \ref{Nonloc} shows that the above expansion indeed provides the dominant behavior of the stationary
velocity distribution as $v \to 0$, although increasingly difficult
to evidence as $a$ increases.

\subsection{Log-normal characteristic approximation}

We now wish to improve upon the heuristic asymptotics \eqref{heurist},
through a more formal derivation.
It proves convenient to redefine $\hat f_{0}(s)$ as the Laplace transform
\beq \hat f_{0}(s)= \int_0^{+\infty} f_{0}(v) e^{-sv} dv\eeq 
which, for imaginary arguments, 
coincides with the previous definition of $\hat f_{0}(s)$ given in section \ref{Colds1}, and 
verifies again
\beq 
\hat f_{0}(s) = \dfrac{\hat f_{0}(sa) }{1+gs} \label{coldlap}
\eeq
Iterating this relation, we obtain
\beq \hat f_{0}\left(\dfrac{1}{g}\right) = \hat f_{0}\left(\dfrac{1}{g a^{n+1}}\right)\,a^{-(n+1)(n+2)/2} \prod_{k=1}^{n+1}   \left(1 + a^k\right) \label{laplead}
\eeq
The product over $k$ may be rewritten as $(-a,a)_{n+1}$ using the $q$-Pochhammer symbol defined in \eqref{defbn}, and it is bounded from above by the finite asymptotic value $(-a,a)_{\infty}$ given as a function of $a$ in \cite{gordon2000some}. We may thus take the limit $n\to \infty$, for which $1/ga^{n+1}\to \infty$. 
However $\hat f_{0}(1/g)$ is finite whereas 
\beq 
\lim_{s\to+\infty} s \hat f_{0}(s) = f_{0}(0) = 0
\eeq 
For $\hat f_{0}(1/g)$ to be finite as $n\to \infty$, we must have
\beq 
\hat f_{0}\left(\dfrac{1}{g a^{n+1}}\right) \propto a^{(n+1)(n+2)/2} \label{asymcond}
\eeq 
Therefore, taking $s=a^{-(n+3/2)}$, we find
\beq
\lim_{s\to \infty} s\, \hat f_{0}(s \sqrt a/g) \propto \exp\left[\frac{\chi \ln^2(s) }{2} \right]
\quad\hbox{with}\quad
\chi = (\ln a)^{-1} <0
\label{asymLap}
\eeq

We emphasize here that in deriving the above asymptotic expansion, everything amounts to
neglecting the loss term in the Boltzmann equation \eqref{coldMax}. 
Indeed, if we expect $f_{0}(v)$ to decrease steeply for 
vanishing velocities, it is much less probable to hold velocity $v\ll 1$ than $v/a$, allowing us to approximate \eqref{coldMax} by 
\beq 
g f_{0}'(v) \approx \dfrac{1}{a} f_{0}\left(\dfrac{v}{a}\right) \label{coldrel}
\eeq
hence for the Laplace transform
\beq 
gs\hat f_{0}(s)\approx \hat f_{0}(as) 
\label{coldlhat}
\eeq
which may be iterated as in \eqref{laplead}, leaving out the factor $(-a,a)_{n+1}$ but giving the same asymptotic condition \eqref{asymcond}.
This equation is verified exactly by the log-normal law $\Psi$:
\beq 
gs \Psi(s) = \Psi(as)\label{lneq}
\eeq 
\beq 
\Psi(s) = \dfrac{1}{s \sqrt{2\pi\sigma}}\exp \left[- \dfrac{(\ln s - m)^2}{2 \sigma^2} \right] 
\eeq 
where $m$ and $\sigma$ are easily determined from the previously known parameters
\beq 
\sigma^2 = - \ln a  > 0 \quad \text{(as }\,\, 0<a<1 \,\, \text{ in this section)}
\eeq 
\beq  
m =  -\dfrac{1}{2}\ln a - \ln g
\eeq  
As discussed in Leipnik \cite{Leipnik} and Lopez \cite{Lopez11}, equation \eqref{lneq} and its counterpart in velocity space \eqref{coldrel} are verified exactly by multiple families of functions. We note in passing that the existence of these different families of solutions is linked with well-known indeterminacy problems surrounding the log-normal: most notable is the non-uniqueness of its moments, { as identical sets of moments may be found in several families of distributions. This problem has been studied since Stieltjes' memoir ``Recherches sur les fractions continues'' \cite{Stieltjes}, which elucidated the fact that the knowledge of the whole set of moments does not determine univocally a probability distribution}.

Assuming that $\hat f_0(s)$ is indeed of log-normal form in the limit $s\to \infty$, we may derive the corresponding asymptotic expression of $f_{0}(v)$ for vanishing velocities, which later receives numerical confirmation. In the spirit of \cite{Leipnik,Bruijn}, this expression is found from an integral expression verifying \eqref{coldrel}
\beq  
\varphi_S(v) =  \int_{k-i\infty}^{k+i\infty} \exp\left[\dfrac{\sigma^2 z^2}{2}-z( \ln v+m)\right] S(z) \Gamma(z) dz \label{charbruijn}
\eeq 
where $S(z)$ is any anti-periodic function of anti-period $1$: $S(z-1)=-S(z)$ such that $\Gamma(z) S(z)$ is analytic for $z\in[k,k+1]$. Leipnik shows 
in \cite{Leipnik} that the choice $S(z) = (2\pi)^{-1} \sin(\pi z)$ -- that we also adopt below -- equates $\varphi_S(v)$ with the characteristic function (i.e. Fourier transform) of the log-normal law. 
Applying Laplace's method to the integral, we expect for large $|\ln v|$
\beq  \varphi_S(v) \approx  \exp\left[\dfrac{\sigma^2 z_0^2}{2}-z_0( \ln v+m) \right] S(z_0) \Gamma(z_0)  \eeq 
with $z_0$ defined as the point where the logarithmic derivative of the integrand vanishes
\beq 
0 = \sigma^2 z_0 -( \ln v+m)  + \dfrac{S'(z_0)}{S(z_0)} + \dfrac{\Gamma'(z_0)}{\Gamma(z_0)}
\eeq 
we may therefore take $z_0 \approx (\ln v +m)/\sigma^2$ as the other terms grow at most logarithmically. Letting $y=\ln v +m$ suggests the asymptotic behavior
\beq
f_{0}\!\left(v=e^{y-m}\right) \propto \exp\left[-\dfrac{y^2}{2\sigma^2} \right] S(y/\sigma^2) \Gamma(y/\sigma^2) \eeq
\beq f_0(v) \propto 
\exp\left[\chi y \left( \dfrac{y}{2} +1 -  \ln(\chi y) \right)\right]  
\quad\hbox{for}\quad 
y \equiv \ln(v /\sqrt{a}g) \to -\infty 
\label{coldsasym}
\eeq
where we used $\chi\ln v \gg 1$ and Leipnik's choice $S^{-1}(z)=2\Gamma(z)\Gamma(1-z)$ \cite{Leipnik}. 
We thus recover the leading order of the heuristic expression  \eqref{heurist}, with additionally
a subleading contribution $(\chi\ln v)^{-\chi\ln v}$ in the probability distribution,
stemming from the gamma function.
Although subleading, this correction is nevertheless important: the complete structure of expression \eqref{coldsasym} is required so that, upon insertion in the Boltzmann equation without loss term, both sides may be matched for vanishing $v$. Indeed, it verifies relation \eqref{coldrel} to leading order in $v/\ln v$ whereas our first log-normal approximation \eqref{heurist} does not.

\begin{figure}[h]
\begin{center}
\includegraphics[height=160 pt, clip]{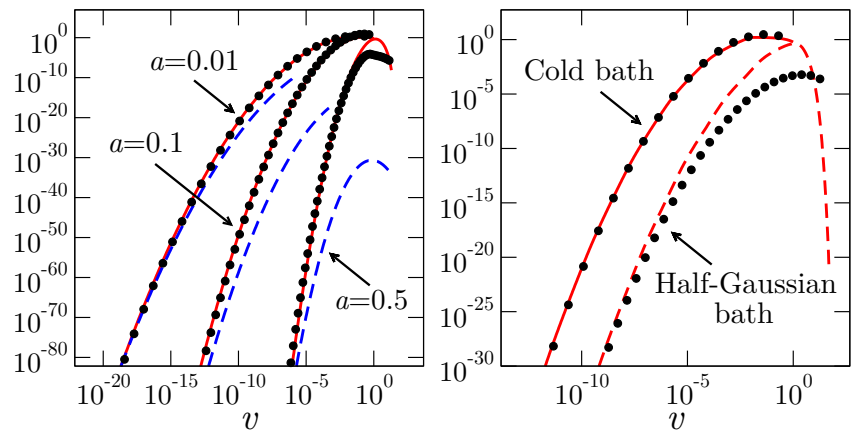}
\caption{Left:  Validity of the asymptotic expressions.
The exact cold bath solution $f_0(v)$ [solid lines, corresponding to
\eqref{coldsol}] against 
the predictions \eqref{heurist} (dashed lines) and \eqref{coldsasym} (dots). 
Here, $a=0.01, 0.1$ and $0.5$ from left to right, and $g=1$. As $a$ increases, the predicted asymptotic behavior appears 
further in the tail, and the expansion \eqref{coldsasym} becomes significantly more precise than our first heuristic guess \eqref{heurist}. 
Right: Close-up on the data for $a=0.01$ (solid line) and comparison with the solution corresponding to a half-Gaussian bath $\Phi(v)=\Theta(v) (2\pi)^{-1/2} \exp(-v^2/2)$ (dashed line), other parameters being identical.
The same asymptotic form \eqref{coldsasym} (dots)
holds in both cases. Similar behavior is expected for any bath distribution with a lower 
velocity bound, provided $g>0$ and $a>0$. }
\label{Nonloc2}
\end{center}
\end{figure}

This new asymptotic expression is clearly seen in Fig. \ref{Nonloc2} (left) to approximate the solution $f_0(v)$ much more closely and on a wider range of velocities than \eqref{heurist}.
Furthermore the right-hand graph demonstrates the same behavior for a different non-negative bath distribution: a half-Gaussian, which strictly vanishes
for $v<0$. This tail behavior appears for $a>0$ as long as there is a lower bound to the bath particle velocity. In addition, any distribution $\Phi(v)\neq \delta(v)$ but bounded from below with support
in $[0,\infty[$ leads to a decrease of
tracer velocity probabilities (see the half-Gaussian results in Fig.
\ref{Nonloc2}, well below their cold bath counterpart), shifting the tail by a multiplicative constant 
without affecting its functional form. However, 
the range of validity of this asymptotic expression is relegated further in the tail when $a$ increases.

\section{Conclusions}
\label{conclusion}

Within a generalized Boltzmann-Lorentz framework, we have studied the stationary
state of a tracer particle immersed in a bath with arbitrary velocity statistics $\Phi(v)$.
The tracer (say a charged particle) is accelerated by an external field,
that does not act on the bath (say made up of neutral particles).
In addition to $\Phi$,
the model is specified by the field intensity $g$, a parameter $a$ accounting for material
properties that combines mass ratio
(tracer over bath particle) together with collisional dissipation, and an index $\nu$ quantifying the
softness of particle interactions. Maxwellian, hard, and very hard rods are thereby embeded
in a unifying approach, that lends itself to analytical
progress and numerical investigation. Three independent simulation techniques
were used to solve the Boltzmann-Lorentz equation.
We have performed an asymptotic analysis of the high-energy tails.
We have evidenced a winner-takes-all competition between the two processes
that leads to a non equilibrium stationary state: 
on the one hand the acceleration during the variable timespan of ballistic progress, which feeds energy into the system, and on the other hand, the collision with scatterers that lead 
to dissipation. Our analysis has revealed that, depending on the parameter and velocity range that 
we consider, either of these effects dominates and completely determines the local shape of 
the distribution. Therefore, significant portions of this distribution are almost independent either 
from the bath or from the acceleration and collision kernel. The latter effect may provide a way to 
probe non-gaussianities in the bath using a more visible tracer and charting out its large velocities 
behavior; Such a behavior requires that the scattering exponent $\nu$ is large
enough once $\Phi$ has been chosen, or conversely that the bath distribution 
is not too cold --i.e. not too peaked around the origin-- once the scattering exponent is
fixed. Whereas most of the analysis was performed for a one dimensional system,
we have, for diffusive properties, also considered driven discs or spheres
in higher dimension.

Motivated by the existence of bath-independent properties, particular emphasis was put
on an already introduced setting, but previously unsolved in its general formulation: 
the "cold gas" with vanishing temperature [i.e. $\Phi(v)=\delta(v)$].
Its interest as an approximation for more realistic models was addressed.
We have in particular shown that even when the cold bath asymptotics does not
prevail, an intermediate asymptotic cold bath regime may exist, when the tracer 
velocity is significantly larger than the bath characteristic velocity but lower than the 
model-dependent threshold $v_c$. This setting also allowed us to exhibit an interesting 
asymptotic behavior at low velocities for a tracer in a bath with a non-negative velocity 
distribution --e.g. a heavy intruder falling in a static bath or along a stationary stream of particles.

\begin{acknowledgements}

We would like to thank A. Alastuey, A. Burdeau, S. Majumdar, J. Talbot, and P. Viot for useful discussions.
\end{acknowledgements}

\bibliography{tracer}

\section*{Appendix: Computation of the diffusion coefficient}
Let us define the velocity autocorrelation function $\Gamma(t)$ as
\beq \Gamma(t) = \left\langle \,\,(v(t)-\left\langle v \right\rangle) \,(v(0)-\left\langle v \right\rangle) \,\,\right\rangle\eeq 
where $\left\langle \,. \,\right\rangle$ denotes the mean over the stationary distribution $f_{\nu}(v)$ and $v(t)$ is the velocity of a given realization of the tracer at time $t$.  A standard relation \cite{Pias83,Resibois} connects $\Gamma(t)$ to the diffusion coefficient {\beqa D& =& \lim_{t\to\infty} \dfrac{1}{2}\dfrac{d}{dt} \left\langle \left( x(t) - \langle v \rangle t\right)^2 \right\rangle \nonumber\\ &= &\lim_{t\to\infty} \left\langle (v(t)-\langle v \rangle) \int_0^t dt' (v(t')-\langle v \rangle) \right\rangle  \nonumber\\
 & =& \int_0^\infty dt\,\Gamma(t)\eeqa where use was made of the fact that both $v(0)$ and $v(\infty)$ are sampled according to $f_\nu(v)$, and it was assumed that the above limits exist.} In a derivation similar to the one used in \cite{Resibois}, this function $\Gamma(t)$ is expressed as
\beq \Gamma(t) =\int dv_0 \,f_{\nu}(v_0) \int d v \, F_{\nu}(v,t|v_0) \,(v-\left\langle v \right\rangle) \,(v_0-\left\langle v \right\rangle)\eeq 
where $F_{\nu}(v,t|v_0)$ is the conditional velocity distribution of the tracer at time $t$ knowing it had velocity $v_0$ at time $0$, i.e. it is the time-dependent distribution $F_{\nu}(v,t)$ with the initial condition $F_{\nu}(v,0) = \delta(v-v_0)$, which may also be represented using the implicit formulation \eqref{implicitF}. We may thus define the auxiliary function
\beq N(v,t) = \int dv_0 \,f_{\nu}(v_0) \, F_{\nu}(v,t|v_0) \,(v_0-\left\langle v \right\rangle) \label{defG}\eeq
which fulfills the initial condition
\beq N(v,0) = (v-\left\langle v \right\rangle) f_{\nu}(v) \label{initialG} \eeq
and
\beq \Gamma(t) = \int d v \,(v-\left\langle v \right\rangle) N(v,t) \label{GammaG}\eeq
Due to the linearity of the Boltzmann-Lorentz equation, $N(v,t)$ follows the same equation as $F_\nu(v,t)$ 
as can be seen from Eq. \eqref{defG}. Therefore, the determination of $\Gamma(t)$ using \eqref{GammaG} amounts to computing the integral and first moment of the solution of the Boltzmann-Lorentz equation with the non-physical initial condition \eqref{initialG}. As this equation conserves the normalization,
\beq \int dv \,N(v,t) = \int dv \,N(v,0) = 0\eeq 
thus
\beq \Gamma(t) = \int dv \,v\,N(v,t)\eeq 
Finally, if $\nu=0$, this time-dependent first moment can be computed directly from the Boltzmann equation
\beq \dfrac{\partial \Gamma}{\partial t} = -  g \int dv \,v \dfrac{\partial N}{\partial v}+ \int dv_1 dv_2 \, N(v_1)\Phi(v_2) \int dv \,v \left[\delta(v_1'-v) - \delta(v_1-v) \right]\label{Geq} \eeq
The first term  on the right-hand side vanishes, while $ v_1' - v_1 = (a-1) \,(v_1 - v_2)$ and therefore
\beq 
\dfrac{\partial \Gamma}{\partial t} = -(1-a)\Gamma(t)
\eeq 
\beq 
\Gamma(t) = \Gamma(0)\,e^{-(1-a)t}
\eeq 
from which the diffusion coefficient follows
\beq D=\dfrac{\Gamma(0)}{1-a}= \dfrac{\left\langle v^2 \right\rangle - \left\langle v \right\rangle^2}{1-a}  \eeq 
We have thus related $D$ to the variance of the tracer stationary velocity distribution
in the case of Maxwell particles with an arbitrary bath. 

Furthermore, the approach can be generalized to higher dimensions, where the collision law reads
 \beq {\bf v'_1} = \mathbf{ v_1} + (1-a) \,[ (\bf v_2-\bf v_1).\hat\sigma ] \hat \sigma\eeq  so that equation \eqref{Geq} becomes
 \beq 
\dfrac{\partial \Gamma}{\partial t} = -   \int d\bv \,\bv \,({\bf g}. \nabla_{\bv}) \,\bG(\bv,t) +\int d\bv_1 d\bv_2 d \hat\sigma\, \bG(\bv_1)\Phi(\bv_2) \int d\bv \,\bv \left[\delta(\bv_1'-\bv) - \delta(\bv_1-\bv) \right]
\label{Geq2} 
\eeq
Then, if we define ${\bf e_{12}}$ the unit vector along $(\bf v_2-\bf v_1)$, $\theta$ the angle between $\hat\sigma$ and ${\bf e_{12}}$, and $\hat\Omega$ the solid angle
\beq 
\int d \hat\sigma [ (\bv_2-\bv_1).\hat\sigma ] \hat \sigma = \int d\hat\Omega |\bv_2-\bv_1| \cos^2(\theta) \,{\bf e_{12}}  =C(d)\,( \bv_2-\bv_1) 
\eeq 
\beq C(d) = \int d\hat\Omega [1-\sin^2(\theta)] = A_d \left[1 - \dfrac{\int_0^\pi d\theta\,  \sin^{d}\theta }{\int_0^\pi d\theta\, \sin^{d-2}\theta  } \right] = \dfrac{A_d}{d}\eeq 
where $A_d$ is the area of the $d$-dimensional unit hypersphere. We may then compute the first and second moments: letting ${\bf g} = g {\bf e_x}$
\beq  \left\langle \bv \right\rangle =  \left\langle v_x \right\rangle {\bf e_x} = \dfrac{{\bf g}d}{A_d(1-a))} \eeq 

\beq  \left\langle \bv^2 \right\rangle = \dfrac{1-a}{1+a} \left\langle v^2 \right\rangle_b + \dfrac{2gd}{1-a^2} \dfrac{ \left\langle v_x \right\rangle }{A_d} \eeq 
hence
\beq \Gamma(0) = \left\langle \bv^2 \right\rangle - \left\langle \bv \right\rangle^2 = \dfrac{1-a}{1+a} \left\langle v^2 \right\rangle_b + \dfrac{g^2d^2}{A_d^2(1-a^2)}  \eeq 
 which gives the following relation for the diffusion coefficient in $d$ dimensions in a bath with unit temperature:

\beq D_d = \dfrac{d}{A_d(1+a)}\left(1 + \dfrac{g^2 d^2}{A_d^2(1-a)^2}\right) \eeq 

\end{document}